\documentclass[fleqn,usenatbib]{mnras}

\usepackage{newtxtext,newtxmath}

\usepackage[T1]{fontenc}

\DeclareRobustCommand{\VAN}[3]{#2}
\let\VANthebibliography\thebibliography
\def\thebibliography{\DeclareRobustCommand{\VAN}[3]{##3}\VANthebibliography}

\usepackage{textcomp,graphicx,amsmath,setspace,multirow,psfrag,color,orcidlink}

\title[Underluminous standardizable candles]{Underluminous 1991bg-like Type Ia supernovae are standardizable candles}
\author[O. Graur]{
O.~Graur\orcidlink{0000-0002-4391-6137}$^{1,2}$\thanks{E-mail: \href{mailto:or.graur@port.ac.uk}{or.graur@port.ac.uk}}
\\
$^1$Institute of Cosmology and Gravitation, University of Portsmouth, Portsmouth, PO1 3FX, UK\\
$^2$Department of Astrophysics, American Museum of Natural History, Central Park West and 79th Street, New York, NY 10024-5192, USA\\
}

\date{Accepted XXX. Received YYY; in original form ZZZ}

\pubyear{2024}

\begin{document}
\label{firstpage}
\pagerange{\pageref{firstpage}--\pageref{lastpage}}
\maketitle



\begin{abstract}
\noindent It is widely accepted that the width-luminosity relation used to standardize normal Type Ia supernovae (SNe Ia) breaks down in underluminous, 1991bg-like SNe Ia. This breakdown may be due to the choice of parameter used as a stand-in for the width of the SN Ia light curve. Using the colour stretch parameter $s_\mathrm{BV}$ instead of older parameters resolves this issue. Here, I assemble a sample of 14 nearby 1991bg-like SNe Ia from the literature, all of which have independent host-galaxy distance moduli and little to no reddening. I use Gaussian process regression to fit the light curves of these SNe in $U/u$, $B$, $V$, $g$, $R/r$, $I/i$, and $H$, and measure their peak absolute magnitudes. I find statistically significant ($>5\sigma$ confidence level in the optical and $>4\sigma$ in the near-infrared) correlations between the peak absolute magnitudes of the 1991bg-like SNe Ia and their $s_\mathrm{BV}$ values in the range $0.2<s_\mathrm{BV}<0.6$. These correlations are broadly consistent with fits to $s_\mathrm{BV}<0.7$ SNe Ia with preliminary $B$- and $V$-band peak absolute magnitudes from the Carnegie Supernova Project and significantly inconsistent with similar fits to normal and transitional SNe Ia (with $0.7<s_\mathrm{BV}<1.1$). The underluminous width-luminosity relation shown here needs to be properly calibrated with a homogeneous sample of 1991bg-like SNe Ia, after which it could be used as a rung on a new cosmological distance ladder. With surface-brightness fluctuations (or another non-Cepheid method) used to calibrate distances to nearby 1991bg-like SNe Ia, such a ladder could produce an independent measurement of the Hubble-Lema\^{i}tre Constant, $H_\mathrm{0}$.

\end{abstract}

\begin{keywords}
methods: distance scale -- supernovae: general 
\end{keywords}


\section{Introduction}
\label{sec:intro}

Type Ia supernovae (SNe Ia) are famous for their use as standard candles, a use that led to the discovery of the accelerating expansion of the Universe and the existence of dark energy \citep{1998AJ....116.1009R,1999ApJ...517..565P}. Yet, SNe Ia are not truly standard; they are \emph{standardizable}. Independently, \citet{1974PhDT.........7R} and \citet{1977SvA....21..675P} both showed that more luminous SNe Ia had wider light curves, i.e., they took longer to rise to peak and then decline. This correlation was corroborated and modernized by \citet{1993ApJ...413L.105P} and has been in use ever since (for a historical review of the width-luminosity relation, see \citealt{2017hsn..book.2543P}; for a general review of the history and physics of SNe Ia, see \citealt{2022supe.book.....G}). 

\citet{1993ApJ...413L.105P} found a tight correlation between the peak absolute luminosities of ten SNe Ia and the widths of their light curves, parameterized by $\Delta m_\mathrm{15}(B)$, the number of magnitudes by which a SN Ia fades 15 days after it peaks in the $B$ band. Variations on this parameter are now routinely used to fit SN Ia light curves and standardize them for use in cosmology. Examples include $\Delta$, used in \texttt{MLCS2k2} \citep{2007ApJ...659..122J}; $s$, which is used in \texttt{SALT} and \texttt{SiFTO} \citep{Guy2005,2008ApJ...681..482C}; and $x1$, used in \texttt{SALT2} \citep{Guy2007}. 

SNe are divided into several classes, each with its own subclasses. The same is true for SNe Ia. The majority of SNe Ia fall into the so-called `normal' subclass, but there are also overluminous 1991T-like SNe Ia \citep{1992ApJ...384L..15F}, transitional 1986G-like SNe Ia \citep{1987PASP...99..592P}, underluminous 1991bg-like SNe Ia \citep{1992AJ....104.1543F}, peculiar SNe Iax \citep{2006AJ....132..189J,2013ApJ...767...57F}, and more. The width-luminosity relation works exceptionally well for normal SNe Ia and extends to overluminous 1991T-like and 1999aa-like SNe Ia as well (e.g.,  \citealt{2018ApJ...869...56B,2022ApJ...938...83Y}). However, this relation breaks down in underluminous, fast-evolving 1991bg-like SNe \citep{2014ApJ...789...32B}. Consequently, these SNe Ia are routinely excised from cosmology samples, and the belief that `1991bg-like SNe Ia are not standard candles' has become dogma. 

\begin{table*}
 \caption{SN sample: basic parameters}\label{table:info}
 \begin{tabular}{lcccccccc}
  \hline
  \hline
  SN & Host & Class$^a$ & Redshift & $\Delta m_\mathrm{15}(B)$ & $s_\mathrm{BV}$ & $\mu^b$ & $E(B-V)_\mathrm{MW}$ & $E(B-V)_\mathrm{Host}$  \\
     &      &           &          & (mag)                     &                 & (mag) & (mag)                & (mag)                   \\ 
  \hline
  \multicolumn{9}{c}{Calibration sample} \\
  \hline
  1991bg  & NGC 4374 & E    & 0.00339 & 1.88(0.10) & 0.321(0.061) & 31.23(0.02) & 0.037 & 0.05(0.02)   \\ 
  1998de  & NGC 0252 & S0   & 0.01647 & 1.95(0.09) & 0.351(0.120) & 33.98(0.20) & 0.052 & 0.00(0.00)   \\ 
  1999by  & NGC 2841 & Sab  & 0.00212 & 1.90(0.05) & 0.381(0.063) & 30.84(0.05) & 0.010 & 0.020(0.030) \\ 
  2005bl  & NGC 4059 & E    & 0.02392 & 2.00(0.06) & 0.387(0.061) & 35.21(0.11) & 0.025 & 0.17(0.08) \\ 
  2006mr  & NGC 1316 & Sab  & 0.00601 & 1.82(0.02) & 0.239(0.060) & 31.42(0.02) & 0.018 & 0.089(0.039) \\ 
  2007ax  & NGC 2577 & S0   & 0.00688 & 1.87(0.06) & 0.355(0.061) & 33.20(0.14) & 0.045 & $<0.01$       \\ 
  2008R   & NGC 1200 & S0   & 0.01350 & 1.82(0.02) & 0.591(0.060) & 33.66(0.08) & 0.062 & 0.009(0.013) \\ 
  2015bo  & NGC 5490 & E    & 0.01614 & 1.91(0.01) & 0.480(0.010) & 34.27(0.08) & 0.023 & 0.00(0.00)   \\ 
  2017ejb & NGC 4696 & E    & 0.00990 & 2.04(0.04) & 0.470(0.010) & 32.88(0.06) & 0.097 & 0.05(0.02)   \\ 
  2019so  & NGC 4622 & Sa   & 0.01457 & 1.96(0.04) & 0.410(0.010) & 34.27(0.48) & 0.128 & -0.01(0.02)  \\ 
  2021qvv & NGC 4442 & S0   & 0.00183 & 2.05(0.03) & 0.280(0.050) & 30.85(0.07) & 0.019 & 0.00(0.00)   \\ 
  \hline
  \multicolumn{9}{c}{Test sample} \\
  \hline
  1997cn  & NGC 5490 & E    & 0.01614 & 1.90(0.05) & 0.350(0.060) & 34.27(0.08) & 0.023 & 0.00(0.00)  \\
  1999da  & NGC 6411 & E    & 0.01269 & 1.94(0.10) & 0.442(0.160) & 33.12(0.27) & 0.046 & 0.00(0.00)  \\
  2022xkq & NGC 1784 & SBc  & 0.00773 & 1.65(0.03) & 0.630(0.030) & 32.14(0.12) & 0.119 & 0.00(0.00)   \\
  \hline
  \multicolumn{9}{l}{$^a$Galaxy classifications sourced from \citet{1991rc3..book.....D}.} \\
  \multicolumn{9}{l}{$^b$All distance moduli are weighted averages of measurements from the literature and independent of distance measurements derived} \\
  \multicolumn{9}{l}{from SN Ia light-curve fitting.}
  \end{tabular}
\end{table*}

\citet{2014ApJ...789...32B} introduced a different stretch parameter, $s_\mathrm{BV}$, defined as the time at which the $B-V$ colour curve reaches peak, relative to the time of $B$-band peak, divided by 30 days. When used as the basis for their own light-curve fitter, \texttt{SNooPy} \citep{2011AJ....141...19B}, this parameter made it possible to standardize not just the light curves of normal and overluminous SNe Ia, but also underluminous SNe Ia as well \citep{2018ApJ...869...56B}. And yet, 1991bg-like SNe Ia are still not treated as standardizable candles. Some of this has to do with the continued use of light-curve fitters such as \texttt{SALT2} and \texttt{MLCS2k2}, which are optimized for normal SNe Ia, and some of it has to do with the now deeply-ingrained belief that 1991bg-like SNe fall off the width-luminosity relation. 

There are hints that the bias against 1991bg-like SNe may be starting to fade. \citet{2018ApJ...869...56B} and \citet{2022ApJ...928..103H} noted that distances derived from the light curves of the 1991bg-like SNe 2006mr and 2015bo were consistent with independent distance measurements of their host galaxies (though see \citealt{2010AJ....140.2036S} and \citealt{2023arXiv231106178D} for dissenting opinions on SN 2006mr). \citet{2018ApJ...869...56B} showed that a quadratic function provided a good fit to absolute magnitudes of both normal and fast-declining SNe Ia across the optical and near-infrared, and that derivations of the Hubble-Lema\^{i}tre constant, $H_\mathrm{0}$, remained consistent whether SNe Ia with $s_\mathrm{BV}>0.5$ were excluded or not. And yet, many cosmological analyses continue to exclude underluminous and fast-declining SNe Ia (e.g., \citealt{2024arXiv240102945V,2024arXiv240102929D}).

In this paper, I set out to independently test whether there is a correlation between the luminosities of 1991bg-like SNe Ia and the widths of their light curves, as parameterized by $s_\mathrm{BV}$. In Section~\ref{sec:data}, I assemble a sample of 14 1991bg-like SNe Ia that suffered little to no dust extinction and that exploded in galaxies with independent distance measurements. Since all SN Ia light-curve fitters are optimized towards normal SNe Ia, in Section~\ref{sec:methods} I fit the light curves of the SNe  myself with Gaussian process regression (GPR). The results, shown in Section~\ref{sec:analysis}, are clear: there are statistically significant correlations between the peak absolute magnitudes of 1991bg-like SNe Ia and $s_\mathrm{BV}$ in multiple filters ($B$, $V$, $g$, $R/r$, $I/i$, and $H$). This leads to Section~\ref{sec:discuss}, in which I discuss the potential use of 1991bg-like SNe Ia in a new distance ladder, independent from the one currently dominating SN Ia cosmology.


\section{Sample}
\label{sec:data}

A review of the literature on 1991bg-like SNe Ia revealed that only $\sim 30$ such objects had been methodically observed and published. Of these, only 14 had a spectroscopic classification, well-sampled multiwavelength light curves, exploded in galaxies with independent distance measurements, and suffered relatively little host-galaxy dust extinction ($E(B-V)<0.2$ mag). The latter is important because most estimates of host-galaxy extinction are a byproduct of light-curve fitting, which I avoid in this paper. The SN sample used in this work is shown in Table~\ref{table:info}. 

Of the 14 SNe, 11 are used to test for the presence of a width-luminosity relation, while the other three SNe are held back as test cases (Section~\ref{subsec:tests}). All of the SNe in this sample were spectroscopically classified as 1991bg-like SNe Ia by various authors, either based on the appearance of the telltale Ti~\textsc{ii} band at 400--440 nm or by comparing their spectra to different SN Ia spectroscopic templates (e.g., using \texttt{SNID}; \citealt{Blondin2007}). Below, I describe the sources of the light curves used in this work. Basic parameters of the SNe in the sample, including the names and types of galaxies in which they exploded, are summarized in Table~\ref{table:info}.

The eponymous SN 1991bg was extensively studied by \citet{1992AJ....104.1543F}, \citet{1993AJ....105..301L}, and \citet{1996MNRAS.283....1T}. Here, I use the peak magnitudes reported by \citet{1996MNRAS.283....1T}, based on a combined analysis of their data and those of \citet{1992AJ....104.1543F} and \citet{1993AJ....105..301L}. $UBVRI$ observations of SN 1997cn were published by \citet{1998AJ....116.2431T} and \citet{2006AJ....131..527J}.  \citet{2001PASP..113..308M} and \citet{2006AJ....131..527J} published $BVRI$ light curves of SN 1998de, obtained as part of the Lick Observatory Supernova Search (LOSS; \citealt{2001ASPC..246..121F}). $UBVRIJHK$ observations of SN 1999by were published by \citet{2004ApJ...613.1120G}, who presented an in-depth study of this object. Additional $BVRI$ observations, obtained by LOSS, were published by \citet{2010ApJS..190..418G}, while three additional epochs of $JHK$ observations were obtained by \citet{2002ApJ...568..791H}. $BVI$ and $BVRI$ observations of SN 1999da were published by \citet{2001AJ....122.1616K} and \citet{2010ApJS..190..418G}, respectively. Here, I use the latter as they contain the additional $R$ filter. 

The Carnegie Supernova Project (CSP; \citealt{2010AJ....139..519C,2011AJ....142..156S,2017AJ....154..211K}) have published $uBVgriYJH$ observations of 16 underluminous, 1991bg-like SNe Ia. Of these, only the host galaxies of SNe 2005bl, 2006mr, 2007ax, and 2008R had independent distance measurements and host-galaxy reddening $<0.2$ mag \citep{2014ApJ...789...32B}. An independent analysis of SN 2005bl \citep{2008MNRAS.385...75T} estimated a host-galaxy reddening of $0.17\pm0.08$ mag and an $R_V$ value of 3.1, which I use here. Additional observations of SN 2007ax were published by \citet{2008ApJ...683L..29K}, who also estimated a negligible host-galaxy reddening of $<0.01$ mag.

SN 2015bo was studied by \citet{2022ApJ...928..103H} in $uBVgriH$. SN 2017ejb was spectroscopically classified as a 1991bg-like SN Ia by \citet{2017ATel10437....1P} and \citet{2017TNSCR.613....1V}. SN 2019so was classified as a 1991bg-like SN Ia by \citet{2019TNSCR..95....1O}. $BVri$ observations of both SNe were obtained with the Las Cumbres Observatory's network of 1-m telescopes \citep{2013PASP..125.1031B} and reported by \citet{2022ApJS..259...53C}. 

Spectra and $UBVgriJH$ images of SN 2021qvv were obtained as part of the Las Cumbres Observatory's Global Supernova Project and published by \citet{2023MNRAS.526.2977G}. The latter classified it as a 1991bg-like SN Ia similar to SN 2006mr. $UBVgriJHK$ observations of SN 2022xkq were published by \citet{2024ApJ...960...29P}, who performed an in-depth analysis of this object.

\section{Analysis}
\label{sec:methods}

Here, I describe the steps taken to measure the peak absolute magnitudes of the SNe in my sample: measuring $s_{\mathrm BV}$ (Section~\ref{subsec:sbv}), obtaining the distance moduli of the SN host galaxies (Section~\ref{subsec:mu}) and the reddening suffered by the SNe (Section~\ref{subsec:red}), and measuring peak apparent and absolute magnitudes (Section~\ref{subsec:peakapp}). 

\subsection{Colour stretch}
\label{subsec:sbv}

\begin{figure}
 \includegraphics[width=0.475\textwidth]{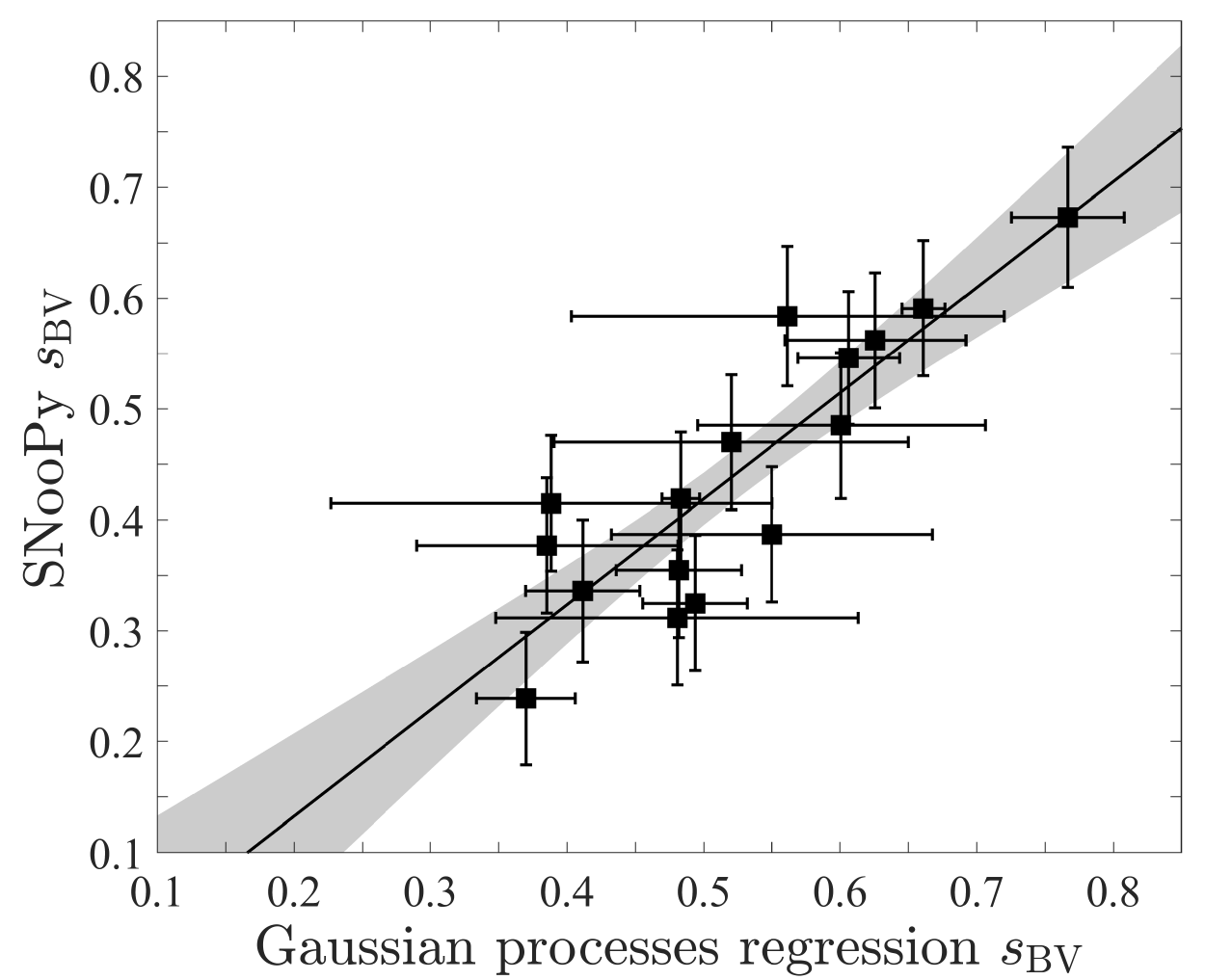}
 \caption{$s_\mathrm{BV}$ values derived with GPR are $\approx 5$ per cent larger than the values derived by the \texttt{SNooPy} light-curve fitter. The correlation shown here is $s_\mathrm{BV}(\mathrm{\texttt{SNooPy}})=(0.95\pm0.14)s_\mathrm{BV}(\mathrm{GPR})-(0.06\pm0.08)$ with a reduced $\chi^2_r$ value of $1.1$.}
 \label{fig:comp}
\end{figure}

SNe 1997cn, 2005bl, 2006mr, 2007ax, 2008R, 2015bo, 2017ejb, 2019so, and 2022xkq have been fit with \texttt{SNooPy} by other authors \citep{2014ApJ...789...32B,2017AJ....154..211K,2022ApJ...928..103H,2022ApJS..259...53C}. For these SNe, I use the published $s_\mathrm{BV}$ values. For the rest of the sample (namely, SNe 1991bg, 1998de, 1999by, 1999da, and 2021qvv), I have derived $s_\mathrm{BV}$ by fitting their light curves with GPR fits, as done by \citet{2023MNRAS.526.2977G}. To ensure that the GPR-derived $s_\mathrm{BV}$ values are not systematically offset from the \texttt{SNooPy}-derived values, I use GPR to fit the light curves of 16 1991bg-like SNe Ia observed by CSP (SNe 2005bl, 2005ke, 2006gt, 2006mr, 2007N, 2007ax, 2007ba, 2008R, 2009F, 2006bd, 2006hb, 2007al, 2007jh, 2008bd, 2008bi, and 2008bt) and compare my values to those published by \citet{2017AJ....154..211K}. As Fig.~\ref{fig:comp} shows, there is a tight correlation between the two techniques, with $s_\mathrm{BV}(\mathrm{\texttt{SNooPy}})=(0.95\pm0.14)s_\mathrm{BV}(\mathrm{GPR})-(0.06\pm0.08)$ and a reduced $\chi^2_r$ value of $15.2/14=1.1$. The GPR-derived $s_\mathrm{BV}$ values of the pre-2000 SNe have been corrected accordingly. The resultant $s_{\mathrm BV}$ values, along with $\Delta m_{15}(B)$ measurements from the literature, are presented in Table~\ref{table:info}.

\subsection{Distance moduli}
\label{subsec:mu}

No single work on nearby distance moduli encompasses all of the host galaxies in my sample. Instead of picking and choosing between the various measurements available for each galaxy, I take a weighted mean of all measurements conducted from the year 2000 onwards. For some host galaxies, which have $>10$ measurements, this results in an underestimated uncertainty on the distance modulus. I do not include distance modulus measurements derived from fitting the light curves of the SNe in the sample, for two reasons. First, this would be cyclical reasoning, as the goal of this paper is to ascertain whether 1991bg-like SNe Ia can be used as standardizable candles. Second, as \citet{2023MNRAS.526.2977G} noted, because current light-curve fitters are not optimized for 1991bg-like SNe Ia, the distance moduli derived in this manner often span a range of $\sim 1$ mag (see Section~\ref{subsec:tests}, below). The sources of the measurements used for each host galaxy are gathered in Appendix~A, and the resultant distance moduli and their uncertainties are collected in Table~\ref{table:info}.

\subsection{Reddening}
\label{subsec:red}

Galactic line-of-sight reddening towards the SNe in the sample, $E(B-V)_\mathrm{MW}$, were extracted from the reddening maps produced by \citet{2011ApJ...737..103S}, while host-galaxy reddening values, $E(B-V)_\mathrm{Host}$, were extracted from the literature (see Section~\ref{sec:data}). Unlike Galactic reddening, host-galaxy reddening is derived from light-curve fits. Since most light-curve fitters are not optimized towards fitting 1991bg-like light curves, these values are suspect. Thus, I chose to include only SNe whose literature $E(B-V)_\mathrm{Host}$ values were $<0.1$ mag. The sole exception is SN 2005bl, with $E(B-V)_\mathrm{Host}=0.17\pm0.08$ mag \citep{2008MNRAS.385...75T}. As it exploded in an elliptical galaxy, I chose to keep it in the sample. To avoid any systematic uncertainties that have to do with how host-galaxy reddening and $R_V$ values are derived, and since, by design, the host-galaxy reddenings are low, I chose not take them into account in the following analysis. A test in which these values were included in the derivation of the peak absolute magnitudes showed that including them did not impact the results in a significant manner.

\subsection{Peak magnitudes}
\label{subsec:peakapp}

Here, I follow \citet{2023MNRAS.526.2977G} and fit the light curves of the SNe in my sample using the Matlab 2023a GPR-fitting routine \texttt{fitrgp} with the default parameters. These fits provide the date and magnitude of maximum light in each of seven filters: $U/u$, $B$, $V$, $g$, $R/r$, $I/i$, and $H$. While some of the SNe in my sample had $YJHK$ observations, most of these observations were obtained after the SNe had already peaked in these bands; the $H$ band was the only one of these filters in which there were more than two SNe with pre- and post-peak observations required for the GPR-fitting routine to estimate peak magnitudes.

To estimate the uncertainties on the derived parameters, I repeat the fit in each filter 100 times, each time varying the photometry of the SN randomly according to the measurement uncertainties. The mean and standard deviation of the results are then taken as the estimated peak date and its $1\sigma$ uncertainty. The same method is applied to estimating the peak apparent magnitude in each filter. Magnitude uncertainties smaller than $0.01$ mag are rounded up. The resultant peak apparent magnitudes are presented in Table~\ref{table:mags}.

\begin{figure}
 \includegraphics[width=0.475\textwidth]{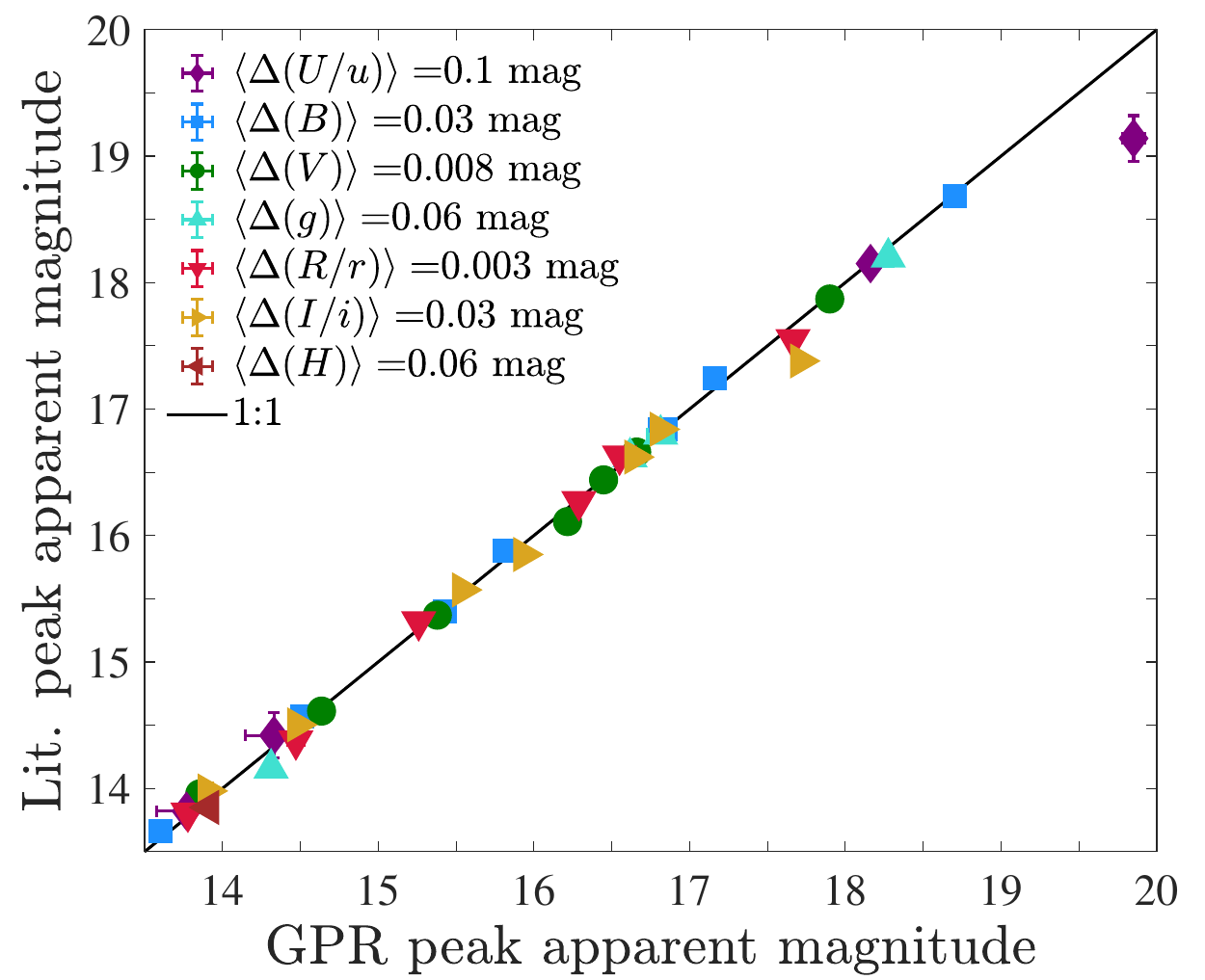}
 \caption{Peak apparent magnitudes derived with GPR are in good agreement with similar values derived by previous works with various light-curve fitters. The solid curve represents the 1:1 line, and the various symbols and colors represent the seven filters tested here (see legend). In six of the seven filters, the mean difference between the GPR and literature values is $\leq0.06$ mag. Only in $U/u$, where the light curves are sparsest, does the mean difference between the values rise to $0.1$ mag, due to a single outlier, SN 2005bl.}
 \label{fig:magcomp}
\end{figure}

In Fig.~\ref{fig:magcomp}, I compare the peak apparent magnitudes derived using this method with published values from the literature, derived using various fitting techniques. In all filters, except $U/u$, the mean difference between the GPR and literature values is $\leq0.06$ mag. In $U/u$, this value rises to $0.1$ mag, but this is due to a single outlier, SN 2005bl.

With the fitted peak apparent magnitudes, Galactic reddening values, and distance moduli described above, the peak absolute magnitude of the $i$-th SN in my sample in filter $\lambda$ is simply:
\begin{equation}\label{eq:abs}
 M(\lambda)_{\mathrm{max},i} = m(\lambda)_{\mathrm{max},i}-A(\lambda)_i-\mu_i,
\end{equation}
where $\mu_i$ is the distance modulus of the $i$-th SN, and $A(\lambda)_i$ is derived from the Galactic line-of-sight reddening, $E(B-V)_{\mathrm{MW},i}$, the Galaxy's $R_V=3.1$, and the \citet{1989ApJ...345..245C} extinction law. Due to the nature of the SN sample, which includes SNe observed by different surveys in different filters (e.g., $U/u$, $R/r$, and $I/i$), I do not attempt to perform $K$ or $S$ corrections. The resultant peak absolute magnitudes are presented in Table~\ref{table:absmags} and Fig.~\ref{fig:MsBV}.


\section{Results}
\label{sec:analysis}

Here, I fit the peak absolute magnitudes measured in the previous section and show that correlations between them and the colour stretch of the SNe are statistically significant in all filters except $U/u$ (Section~\ref{subsec:correlations}); compare these correlations to fits applied to independent CSP data (Section~\ref{subsec:comparison}); and demonstrate the effect of these correlations on our understanding of SNe 1997cn, 1999da, and 2022xkq (Section~\ref{subsec:tests}).

\subsection{Width-luminosity relations}
\label{subsec:correlations}

I perform two independent tests to determine whether the peak absolute magnitudes in a specific filter are correlated with colour stretch, $s_\mathrm{BV}$, the results of which are collected in Table~\ref{table:params}. First, all filters display a strong, negative Pearson correlation coefficient, $r$, with $r\lesssim-0.9$ in $B$, $V$, $g$, $R/r$, and $H$, and $r\sim-0.7$ in $U/u$ and $I/i$. The same test shows that these correlations have statistically significant $p$-values of $<0.05$ in all filters except $U/u$ (and a more stringent $p<0.01$ in all filters except $U/u$ and $I/i$). Second, a likelihood-ratio test (see, e.g., \citealt{2017ApJ...837..120G}) prefers a 1st-order polynomial over a 0th-order polynomial fit to the data in every filter with a statistical significance of $>5\sigma$. The same test shows that a 2nd-order polynomial is not preferred over a 1st-order polynomial.

\begin{figure*}
 \includegraphics[width=0.975\textwidth]{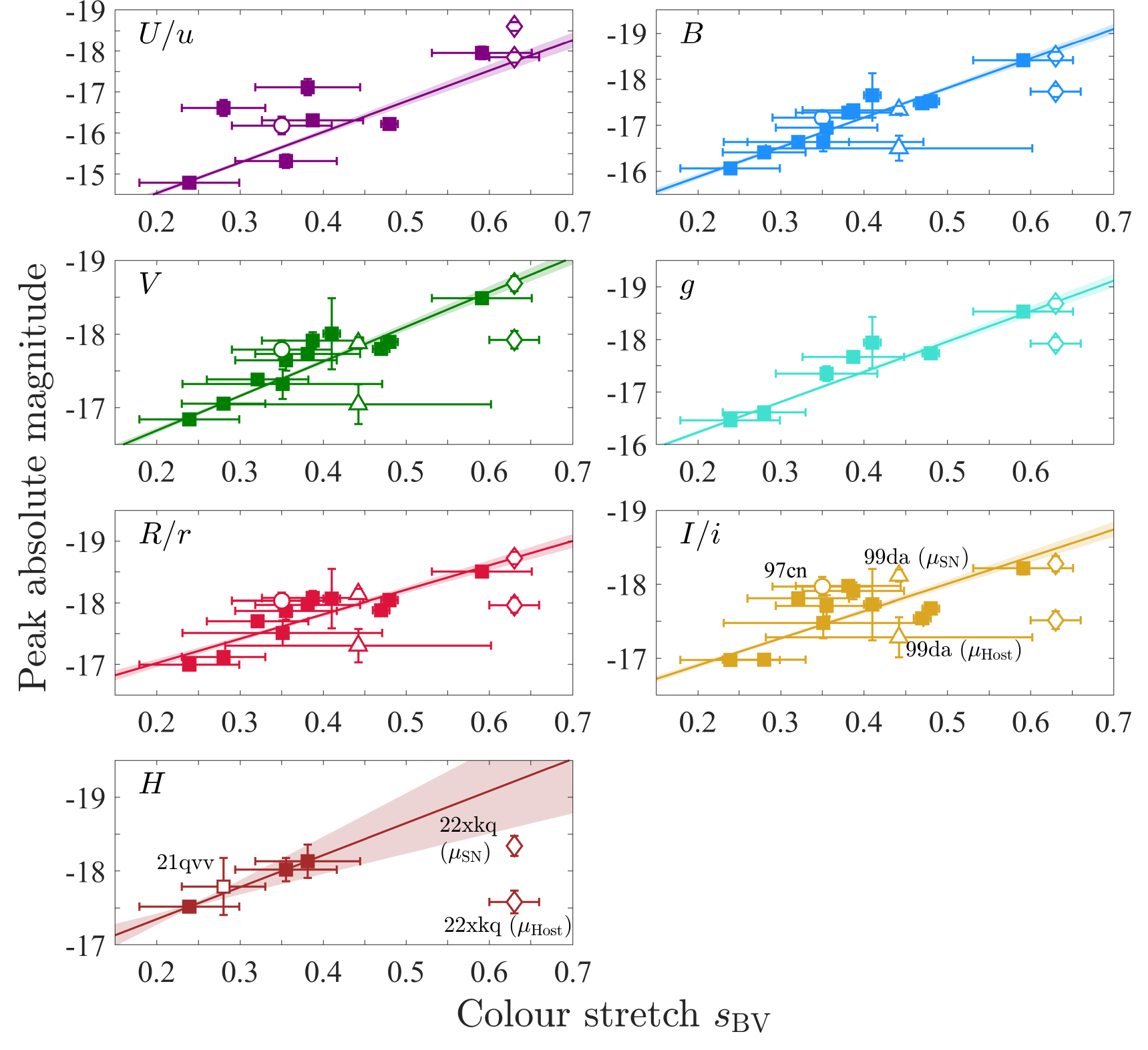}
 \caption{Peak absolute magnitude vs. colour stretch, $s_\mathrm{BV}$. Magnitudes have been corrected for Galactic line-of-sight reddening. The shaded regions around the linear fits represent the $1\sigma$ uncertainties of the fits. The correlations, calculated using the calibration sample (filled squares) are statistically significant in all filters except $U/u$. SNe 1997cn, 1999da, and 2022xkq (described in detail in Section~\ref{subsec:tests}) are shown as an open circle, triangle, and diamond, respectively. SNe 1999da and 2022xkq are shown twice, once when using a host-based distance modulus ($\mu_\mathrm{Host}$) and once when using a distance modulus derived by SN light-curve fitters ($\mu_\mathrm{SN}$). For clarity, the second measurements of SNe 1999da and 2022xkq are shown without their horizontal error bars. The $H$-band measurement of SN 2021qvv, shown as an open square, is not used in the fit.}
 \label{fig:MsBV}
\end{figure*}

Based on the likelihood-ratio test, I fit 1st-order polynomials to the peak absolute magnitudes in each filter, the results of which are presented in Table~\ref{table:params} and Fig.~\ref{fig:MsBV}. An analysis of the residuals produced by subtracting the best-fitting polynomials from the data reveals a scatter, $S$, of $<0.3$ mag in all filters except $U/u$, and $<0.2$ mag in $B$, $V$, and $H$. I attribute the high scatter in $U/u$ to the difficulty in estimating the time of peak from photometry that, only in this band, did not always cover the rise and peak of the light curves (SNe 1999by, 2007ax, 2008R, and 2021qvv). Note that, even without $K$ and $S$ corrections, and with distance moduli derived using a variety of techniques, the scatter found here is on par with, or better than, the scatter found in the initial width-luminosity relations measured by \citet{1993ApJ...413L.105P} for normal SNe Ia.

It is instructive to note that the width-luminosity relation in the $H$ band is similar in nature to the optical correlations. This is not the case in normal SNe Ia, which exhibit a smaller scatter in luminosity in the NIR than in the optical (e.g., \citealt{2019ApJ...887..106A}). However, a larger NIR sample is required to validate this statement, as the quadratic fits shown in figure 4 of \citet{2018ApJ...869...56B} hint at a reduction in the NIR luminosity scatter similar to that seen in normal SNe Ia.

\begin{table*}
 \caption{Peak apparent magnitudes}\label{table:mags}
 \begin{tabular}{lccccccc}
  \hline
  \hline
  SN & $m(U/u)_\mathrm{max}$ & $m(B)_\mathrm{max}$ & $m(V)_\mathrm{max}$ & $m(g)_\mathrm{max}$ & $m(R/r)_\mathrm{max}$ & $m(I/i)_\mathrm{max}$ & $m(H)_\mathrm{max}$ \\
     & (mag)        & (mag)        & (mag)        & (mag)        & (mag)        & (mag) & (mag)        \\
  \hline
  \multicolumn{8}{c}{Calibration sample} \\
  \hline
  1991bg  & $\cdots$    & 14.75(0.10) & 13.96(0.05) & $\cdots$    & 13.63(0.05) & 13.50(0.05) & $\cdots$ \\
  1998de  & $\cdots$    & 17.56(0.04) & 16.82(0.04) & $\cdots$    & 16.61(0.04) & 16.61(0.05) & $\cdots$ \\ 
  1999by  & 13.77(0.19) & 13.60(0.01) & 13.14(0.01) & $\cdots$    & 12.90(0.01) & 12.89(0.01) & 12.71(0.22) \\ 
  2005bl  & 19.85(0.07) & 18.70(0.02) & 17.90(0.02) & 18.28(0.02) & 17.66(0.03) & 17.71(0.03) & $\cdots$ \\
  2006mr  & 16.72(0.03) & 15.43(0.04) & 14.64(0.01) & 15.03(0.01) & 14.47(0.05) & 14.48(0.01) & 13.91(0.01) \\ 
  2007ax  & 18.10(0.09) & 16.44(0.02) & 15.69(0.02) & 16.02(0.03) & 15.45(0.01) & 15.59(0.01) & 15.21(0.07) \\ 
  2008R   & 16.01(0.13) & 15.51(0.01) & 15.37(0.01) & 15.37(0.01) & 15.32(0.01) & 15.58(0.02) & $\cdots$ \\ 
  2015bo  & 18.16(0.04) & 16.85(0.01) & 16.45(0.01) & 16.62(0.01) & 16.29(0.01) & 16.65(0.02) & $\cdots$ \\ 
  2017ejb & $\cdots$    & 15.81(0.02) & 15.38(0.02) & $\cdots$    & 15.26(0.02) & 15.54(0.02) & $\cdots$ \\ 
  2019so  & $\cdots$    & 17.16(0.03) & 16.66(0.02) & 16.81(0.08) & 16.55(0.02) & 16.81(0.04) & $\cdots$ \\ 
  2021qvv & 14.33(0.19) & 14.51(0.01) & 13.85(0.01) & 14.31(0.03) & 13.78(0.01) & 13.91(0.02) & 13.07(0.38)$^a$ \\ 
  
 \hline
 \multicolumn{8}{c}{Test sample} \\
  1997cn (GPR)        & 18.20(0.20) & 17.20(0.10) & 16.55(0.10) & $\cdots$    & 16.30(0.10) & 16.35(0.10) & $\cdots$ \\
  1997cn (2007ax)     & 18.23(0.20) & 17.16(0.10) & 16.41(0.10) & 16.74(0.03) & 16.17(0.10) & 16.31(0.10) & $\cdots$ \\ 
  1999da              & $\cdots$    & 16.81(0.03) & 16.22(0.01) & $\cdots$    & 15.94(0.02) & 15.93(0.02) & $\cdots$ \\
  2022xkq             & 14.87(0.02) & 14.91(0.01) & 14.58(0.01) & 14.67(0.01) & 14.50(0.01) & 14.87(0.01) & 14.63(0.09) \\
 \hline
 \multicolumn{8}{l}{$^a$The $H$-band peak magnitude of SN 2021qvv was derived by \citet{2023MNRAS.526.2977G} using the \texttt{SNooPy} light-curve fitter and a single} \\
 \multicolumn{8}{l}{$H$-band measurement of $13.16 \pm 0.06$ mag obtained $0.6$ d past $B$-band maximum light (see Section~\ref{subsubsec:2021qvv}).}
 \end{tabular}
\end{table*}

\begin{table*}
 \caption{Peak absolute magnitudes}\label{table:absmags}
 \begin{tabular}{lccccccc}
  \hline
  \hline
  SN & $M(U/u)_\mathrm{max}$ & $M(B)_\mathrm{max}$ & $M(V)_\mathrm{max}$ & $M(g)_\mathrm{max}$ & $M(R/r)_\mathrm{max}$ & $M(I/i)_\mathrm{max}$ & $M(H)_\mathrm{max}$ \\
     & (mag)        & (mag)        & (mag)        & (mag)        & (mag)        & (mag) & (mag)       \\
  \hline
  \multicolumn{8}{c}{Calibration sample} \\
  \hline
  1991bg  & $\cdots$     & -16.64(0.10) & -17.38(0.05) & $\cdots$     & -17.70(0.05) & -17.81(0.05) & $\cdots$ \\ 
  1998de  & $\cdots$     & -16.64(0.20) & -17.32(0.20) & $\cdots$     & -17.51(0.20) & -17.48(0.21) & $\cdots$ \\ 
  1999by  & -17.12(0.20) & -17.28(0.05) & -17.73(0.05) & $\cdots$     & -17.97(0.05) & -17.97(0.05) & -18.13(0.23) \\ 
  2005bl  & -16.31(0.13) & -17.33(0.11) & -17.91(0.11) & -17.67(0.11) & -18.07(0.11) & -17.91(0.11) & $\cdots$ \\ 
  2006mr  & -14.79(0.04) & -16.07(0.04) & -16.84(0.02) & -16.46(0.02) & -17.00(0.06) & -16.98(0.02) & -17.52(0.02) \\ 
  2007ax  & -15.32(0.17) & -16.95(0.14) & -17.65(0.14) & -17.35(0.14) & -17.87(0.14) & -17.71(0.14) & -18.02(0.16) \\ 
  2008R   & -17.95(0.13) & -18.41(0.08) & -18.49(0.08) & -18.53(0.08) & -18.51(0.08) & -18.21(0.08) & $\cdots$ \\ 
  2015bo  & -17.52(0.08) & -16.22(0.09) & -17.74(0.08) & -17.89(0.08) & -18.04(0.08) & -17.67(0.08) & $\cdots$ \\ 
  2017ejb & $\cdots$     & -17.48(0.06) & -17.80(0.06) & $\cdots$     & -17.88(0.06) & -17.54(0.06) & $\cdots$ \\ 
  2019so  & $\cdots$     & -17.65(0.48) & -18.01(0.48) & -17.94(0.49) & -18.07(0.48) & -17.73(0.48) & $\cdots$ \\ 
  2021qvv & -16.61(0.20) & -16.42(0.07) & -17.05(0.07) & -16.61(0.07) & -17.12(0.07) & -16.98(0.07) & -17.79(0.39)$^a$ \\ 
 \hline
 \multicolumn{8}{c}{Test sample} \\
  1997cn (GPR)        & -16.18(0.22) & -17.17(0.13) & -17.79(0.13) & $\cdots$     & -18.03(0.13) & -17.97(0.13) & $\cdots$ \\
  1997cn (2007ax)     & -16.16(0.22) & -17.21(0.13) & -17.93(0.13) & -17.62(0.09) & -18.16(0.13) & -18.01(0.13) & $\cdots$ \\ 
  1999da ($\mu_\mathrm{Host}$)  & $\cdots$ & -16.51(0.27) & -17.05(0.27) & $\cdots$ & -17.31(0.27) & -17.28(0.27) & $\cdots$ \\
  1999da ($\mu_\mathrm{SN}$)    & $\cdots$ & -17.34(0.08) & -17.88(0.08) & $\cdots$ & -18.14(0.08) & -18.12(0.08) & $\cdots$ \\ 
  2022xkq ($\mu_\mathrm{Host}$) & -17.84(0.12) & -17.73(0.12) & -17.92(0.12) & -17.92(0.12) & -17.96(0.12) & -17.52(0.12) & -17.58(0.15) \\
  2022xkq ($\mu_\mathrm{SN}$)   & -18.60(0.10) & -18.49(0.10) & -18.68(0.10) & -18.68(0.10) & -18.72(0.10) & -18.28(0.10) & -18.34(0.14)\\
 \hline
 \multicolumn{8}{l}{$^a$The $H$-band peak magnitude of SN 2021qvv was derived by \citet{2023MNRAS.526.2977G} using the \texttt{SNooPy} light-curve fitter and a single} \\
 \multicolumn{8}{l}{$H$-band measurement of $13.16 \pm 0.06$ mag obtained $0.6$ d past $B$-band maximum light (see Section~\ref{subsubsec:2021qvv}).}
 \end{tabular}
\end{table*}

\begin{table*}
 \centering
 \caption{$M$--$s_\mathrm{BV}$ correlation parameters}\label{table:params}
 \begin{tabular}{lccccccc}
  \hline
  \hline
  Filter & $a$   & $b$   & $\chi^2_r$ & $L$        & $r$  & $p$ & $S$   \\
         & (mag) & (mag) &            & ($\sigma$) &      &     & (mag) \\
  \hline
  
  $U/u$ & -7.5(0.2) & -13.0(0.1) & 26  & 5 & -0.72 & $6.6\times10^{-2}$ & 0.74 \\
  
  $B$ & -6.4(0.2) & -14.6(0.1) & 3.9 & 5 & -0.96 & $4.1\times10^{-6}$ & 0.19 \\
  
  $V$ & -4.7(0.2) & -15.8(0.1) & 3.9 & 5 & -0.93 & $4.4\times10^{-5}$ & 0.18 \\
  
  $g$ & -5.8(0.2) & -15.1(0.1) & 1.6 & 5 & -0.95 & $1.1\times10^{-3}$ & 0.23 \\
  
  $R/r$ & -4.0(0.2) & -16.2(0.1) & 7.9 & 5 & -0.88 & $3.1\times10^{-4}$ & 0.21 \\
  
  $I/i$ & -3.7(0.2) & -16.2(0.1) & 23  & 5 & -0.72 & $1.2\times10^{-2}$ & 0.28 \\
  
  $H$   & -4.3(1.0) & -16.5(0.3) & $7\times10^{-4}$ & 4 & -1.0 & $2.6\times10^{-3}$ & $1.4\times10^{-3}$ \\
  
  \hline
  \multicolumn{8}{l}{Columns are: (1) filter; ($2+3$) best-fitting parameters for a first-order polynomial} \\
  \multicolumn{8}{l}{$M=as_\mathrm{BV}+b$ (uncertainties appear in parentheses); (4) reduced $\chi^2$ of fit;} \\
  \multicolumn{8}{l}{(5) significance of likelihood-ratio test; ($6+7$) Pearson $r$ coefficient and its $p$-value;} \\
  \multicolumn{8}{l}{and (8) scatter after subtracting the best-fitting polynomial from the data.} 
 \end{tabular}
\end{table*}

\begin{figure*}
 \begin{tabular}{cc}
  \includegraphics[width=0.475\textwidth]{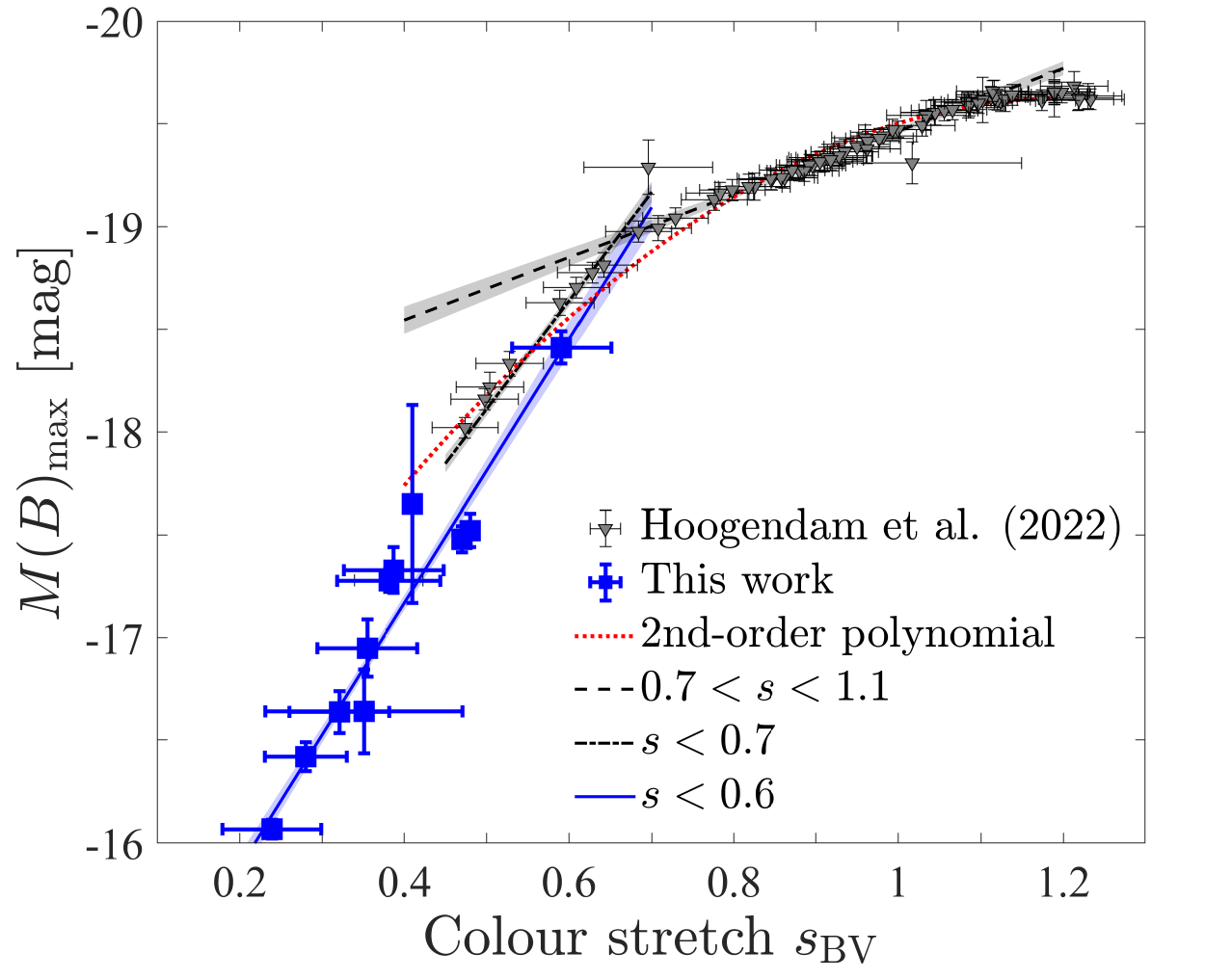} &
  \includegraphics[width=0.475\textwidth]{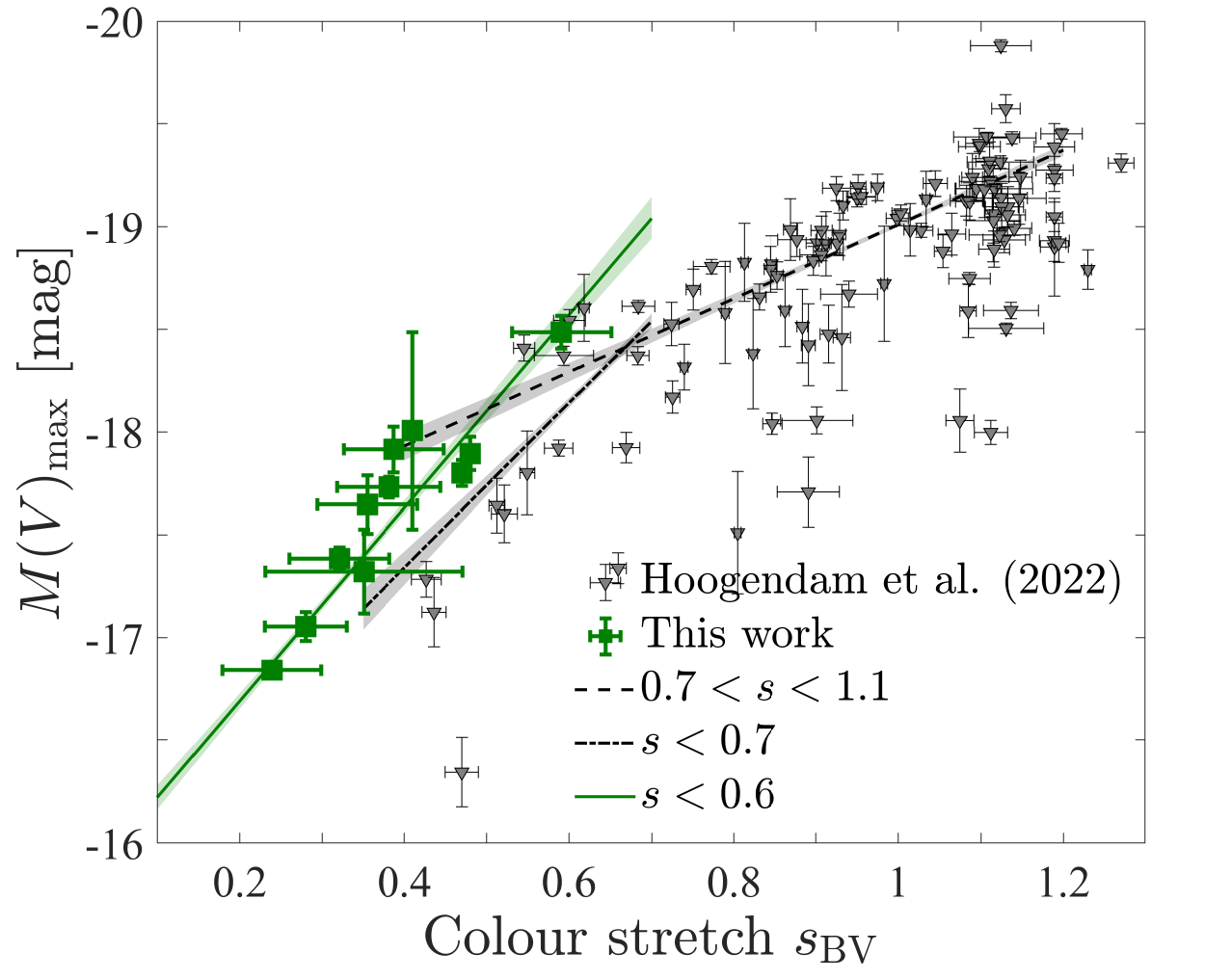} \\
 \end{tabular}
 \caption{Absolute $B$- (left) and $V$-band (right) magnitudes vs. colour stretch, $s_\mathrm{BV}$ for the 1991bg-like SNe Ia used to calibrate the width-luminosity relations in this work (blue and green squares, respectively) and CSP SNe Ia (gray triangles; preliminary measurements shown in \citealt{2022ApJ...928..103H}, with finalised measurements expected in \citealt{2023arXiv230801875U}). The dashed, black curves represent 1st-order polynomial fits to normal and transitional SNe Ia ($0.7<s_\mathrm{BV}<1.1$), while the dot-dashed, black curves were fit to CSP SNe Ia with $s_\mathrm{BV}<0.7$. The red, dotted curve is a 2nd-order polynomial fit to all of the CSP data. As detailed in the text, the normal and underluminous SN Ia curves have significantly different slopes. On the other hand, the slopes of the fits to the peak absolute magnitudes measured here (blue and green solid curves, respectively) are broadly consistent with those of the underluminous CSP SNe Ia.}
 \label{fig:hoogendam}
\end{figure*}

\subsection{Comparison with CSP}
\label{subsec:comparison}

In Fig.~\ref{fig:hoogendam}, I compare my $B$- and $V$-band measurements to similar CSP peak absolute magnitudes from \citet{2022ApJ...928..103H}. These measurements are preliminary (the finalized CSP measurements will appear in the published version of \citealt{2023arXiv230801875U}; W.~B. Hoogendam, private communication) but provide a good sanity test and the option to compare the correlations found above with the well-established width-luminosity relation of normal SNe Ia.

The $B$-band CSP data were obtained by fitting the light curves with \texttt{SNooPy} using the EBV2 model, while the $V$-band peak absolute magnitudes were obtained via GPR fits, which explains the larger scatter in this filter \citep{2022ApJ...928..103H}. To recreate figure 9 from \citet{2022ApJ...928..103H}, I have removed all $V$-band measurements with uncertainties $\sigma[M(V)]>0.3$ mag and $\sigma(s_\mathrm{BV})>0.06$ (\citealt{2022ApJ...928..103H} used a stricter $\sigma(s_\mathrm{BV})>0.05$, but my sample has an average $\langle \sigma(s_\mathrm{BV})\rangle\sim0.06$). No cuts were applied to the $B$-band data.

A likelihood-ratio test applied to the entire range of CSP data in the $B$ band prefers a 2nd-order polynomial over a 1st-order polynomial at a $>5\sigma$ confidence level, indicating that the underluminous SNe Ia follow a different trend than the normal ones. This is consistent with \citet{2018ApJ...869...56B}. However, a visual inspection of the 2nd-order polynomial fit, shown in Fig.~\ref{fig:hoogendam}, reveals that it underestimates the absolute magnitudes in the range $0.55\lesssim s_\mathrm{BV} \lesssim 0.75$. In other words, it does not manage to recreate the sharp break between normal and underluminous SNe Ia. The same likelihood-ratio test conducted on the $V$-band CSP measurements finds no statistically significant preference for a 2nd-order polynomial over a 1st-order polynomial, or even for a 1st-order polynomial over a 0th-order polynomial. 

Based on the results of the likelihood-ratio test on the $B$-band CSP data, I fit the CSP measurements in both filters with 1st-order polynomials in two $s_\mathrm{BV}$ ranges: $0.7<s_\mathrm{BV}<1.1$ to capture all normal SNe Ia (as well as transitional SNe Ia), and $s_\mathrm{BV}<0.7$ to include only underluminous SNe Ia. In the $B$ band, I find that the normal SNe Ia follow $M(B)_\mathrm{CSP}= (-1.5 \pm 0.1)s_\mathrm{BV} -(17.9\pm0.1)~(\chi^2=5.7/42=0.14)$ and underluminous SNe Ia follow $M(B)_\mathrm{CSP}= (-5.3 \pm 0.2)s_\mathrm{BV} -(15.5\pm0.1)~(\chi^2=22/9=2.4)$. In the $V$ band, the correlations are $M(V)_\mathrm{CSP}= (-1.8 \pm 0.1)s_\mathrm{BV} -(17.2\pm0.1)~(\chi^2=1010/55=18)$ and $M(V)_\mathrm{CSP}= (-4.0 \pm 0.2)s_\mathrm{BV} -(15.7\pm0.1)~(\chi^2=498/15=33)$, respectively. The higher $\chi^2$ values, divided by the number of degrees of freedom, are due to the larger scatter of the measurements in this filter.

Overplotting the $B$- and $V$-band peak absolute magnitudes from Fig.~\ref{fig:MsBV} shows that the width-luminosity relations from my sample are in broadly good agreement with the fits to the CSP data. This is somewhat surprising, given the heterogeneity of my sample, the variety of methods used to derive distance moduli to the SN host galaxies, and the lack of $K$ corrections. It is too much to assume that all of these factors would conspire to produce correlations that coincidentally agreed with the homogeneous, well-calibrated CSP data. Instead, we must conclude that there are, indeed, clear correlations between the peak absolute magnitudes of 1991bg-like SNe Ia and the widths of their light curves, as parameterized by $s_\mathrm{BV}$.

\subsection{Test cases}
\label{subsec:tests}

When constructing the SN sample in Section~\ref{sec:data}, the underluminous SNe 1997cn and 1999da were set aside. Below, I describe the circumstances that led me to isolate each of these SNe and how they help validate the width-luminosity relations from Section~\ref{subsec:correlations}. I also discuss the $H$-band peak absolute magnitude of SN 2021qvv as well as the absolute magnitudes of SN 2022xkq, whose situation is similar to that of SN 1999da.
                       
\subsubsection{SN 1997cn}
\label{subsubsec:1997cn}

SN 1997cn was observed by both \citet{1998AJ....116.2431T} and \citet{2006AJ....131..527J}, but only after it had already peaked and begun to decline. \citet{1998AJ....116.2431T} estimated that the SN reached $B$-band maximum light on JD 2450587.5, while \citet{2022ApJ...928..103H} used \texttt{SNooPy} to fit the \citet{2006AJ....131..527J} light curves and derive a peak date of $2450580.7 \pm 0.7$, 6.8 d earlier than the \citet{1998AJ....116.2431T} date and $\sim 2$ d prior to its discovery \citep{1997IAUC.6661....1L}. A spectroscopic fit with \texttt{SNID} yielded a third date: 2450588.3, 0.8 d after the \citet{1998AJ....116.2431T} estimate. As the light curves of this SN did not cover the date of maximum light, I did not include it in my sample.

\begin{figure}
 \includegraphics[width=0.475\textwidth]{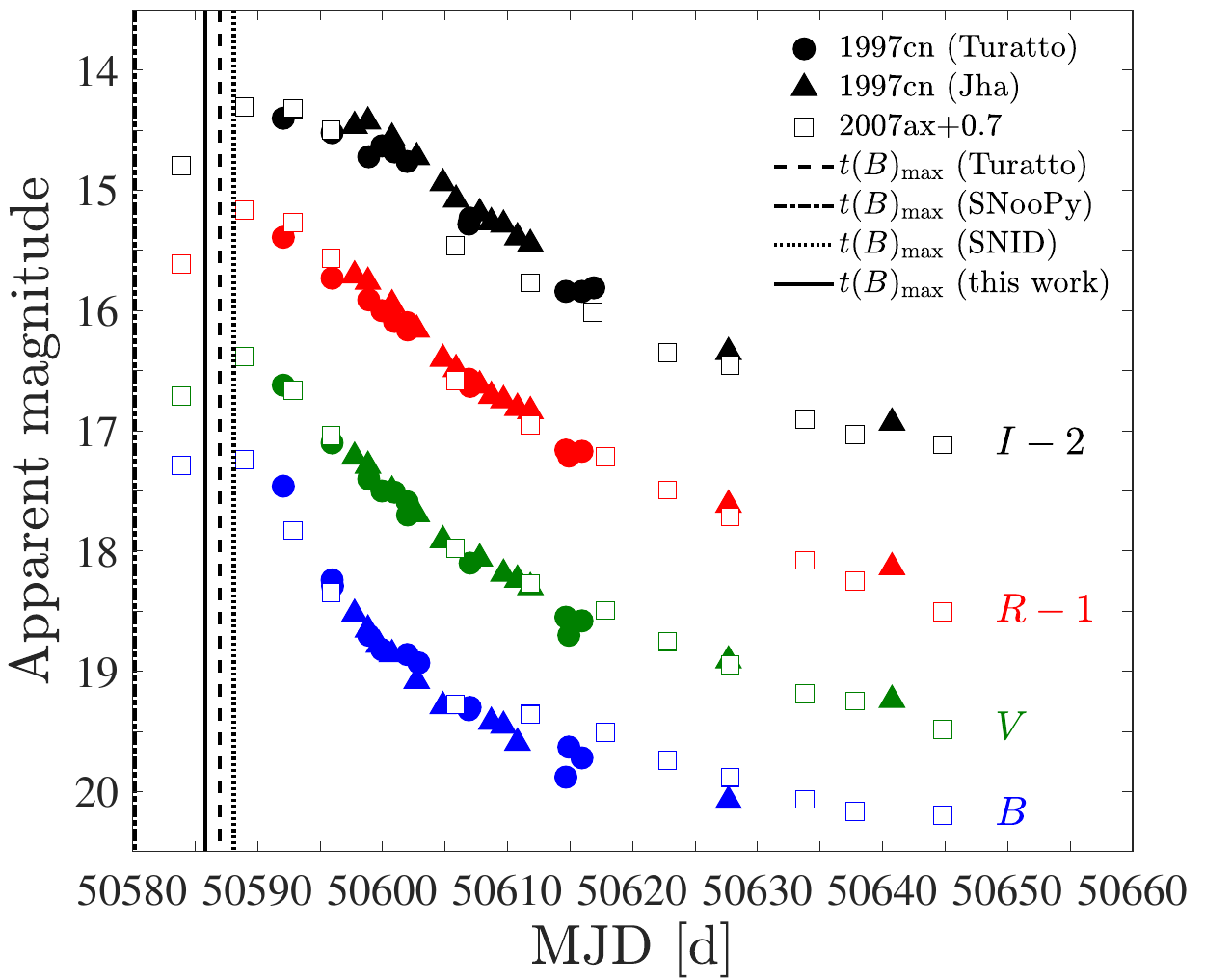}
 \caption{A comparison between SN 1997cn (filled circles and triangles from \citealt{1998AJ....116.2431T} and \citealt{2006AJ....131..527J}, respectively) and SN 2007ax (empty squares; \citealt{2017AJ....154..211K}). Derived dates of maximum light are shown as dashed \citep{1998AJ....116.2431T}, dot-dashed (\texttt{SNooPy}; \citealt{2022ApJ...928..103H}), dotted (\texttt{SNID}; \citealt{2022ApJ...928..103H}), and solid (this work) curves. According to the fit between the two SNe, SN 1997cn was $\approx 0.7$ mag dimmer than SN 2007ax.}
 \label{fig:97cn}
\end{figure}

The \texttt{SNooPy} fit conducted by \citet{2022ApJ...928..103H} provided an $s_\mathrm{BV}$ value of $0.35\pm0.06$. As this value is consistent with the $s_\mathrm{BV}$ value of SN 2007ax, $0.355\pm0.061$ \citep{2017AJ....154..211K}, the light curves of the two SNe can be compared to each other directly. In Fig.~\ref{fig:97cn}, I fit the $V$-band measurements of SN 2007ax to those of SN 1997cn taken by both \citet{1998AJ....116.2431T} and \citet{2006AJ....131..527J} by varying the date of maximum light and the offset between the light curves. I find $t(B)_\mathrm{max} = 2450586.1 \pm 0.2$ d, with SN 1997cn $0.72\pm0.01$ mag fainter than SN 2007ax, with a reduced $\chi^2_r=2.1$. 

Fig.~\ref{fig:MsBV} includes peak absolute magnitudes of SN 1997cn estimated by adding an offset of $0.72$ mag to the peak apparent magnitudes of SN 2007ax. The resultant values are consistent with peak apparent magnitudes estimated by GPR fits to the data from \citet{2006AJ....131..527J}, with the exception of the $g$ band, in which there were no previous measurements. In all filters, the resultant peak absolute magnitudes of SN 1997cn are consistent with the width-luminosity relations within the scatter of the calibration sample. This indicates that the width-luminosity relations measured in Section~\ref{subsec:correlations} can be used to estimate the peak absolute magnitudes of individual 1991bg-like SNe Ia.

\subsubsection{SN 1999da}
\label{subsubsec:1999da}

The host galaxy of SN 1999da only had two independent, post-2000 distance measurements (both from \citealt{2001MNRAS.327.1004B}), which averaged to $\mu_\mathrm{Host} = 33.12 \pm 0.27$ mag. However, distance moduli derived by fitting the light curves of this SN by various groups \citep{Riess2004,2006ApJ...645..488W,2007ApJ...659..122J,2008ApJ...686..749K,2009ApJ...700.1097H}, using different light-curve fitters, produced an average value of $\mu_\mathrm{SN} = 33.95 \pm 0.08$, $\sim 0.8$ mag fainter. Fig.~\ref{fig:MsBV} shows SN 1999da twice, using each of the distance moduli above. In all four filters in which this SN was observed, the peak absolute magnitudes derived using the host-based distance modulus are systematically dimmer than the width-luminosity relations. The peak absolute magnitudes derived with the light-curve-based distance modulus, on the other hand, are consistent with the correlations. This indicates that, at least for this SN, existing light-curve fitters already did a good job of deriving its luminosity distance.

\subsubsection{SN 2021qvv}
\label{subsubsec:2021qvv}

Only one epoch of $JH$ observations was obtained for SN 2021qvv, $0.6$ d after $B$-band maximum light. A \texttt{SNooPy} fit determined that the SN peaked in the $H$ band $\sim 4$ d later, and that the peak magnitude was $13.07 \pm 0.38$ mag, $\sim0.9$ mag brigther than the sole $H$-band measurement \citep{2023MNRAS.526.2977G}. I did not use this light-curve-fitter derived magnitude to fit the correlation between the $H$-band absolute magnitudes and $s_\mathrm{BV}$, yet, when plotted in Fig.~\ref{fig:MsBV}, it is fully consistent with the $H$-band correlation.

\subsubsection{SN 2022xkq}
\label{subsubsec:20222xkq}

The host galaxy of SN 2022xkq has several distance modulus measurements (Table~\ref{table:dm}), which provide a weighted average of $\mu_\mathrm{Host} = 32.14 \pm 0.12$ mag. However, as in the case of SN 1999da, a \texttt{SNooPy} fit provides a $\sim 0.8$ mag fainter distance modulus of $\mu_\mathrm{SN} = 32.9 \pm 0.1$ mag \citep{2024ApJ...960...29P}. Once again, I use both distance moduli to plot the absolute magnitudes of this SN in Fig.~\ref{fig:MsBV}. Exactly as before, the absolute magnitudes derived using the light-curve-based distance modulus are consistent with the correlations (in the $U/u$ band, both distance moduli result in absolute magnitudes that are consistent with the scatter around the correlation, while in the $H$ band, the light-curve-based absolute magnitude is consistent with the correlation within $2\sigma$). This provides another indication that light-curve fitters, and especially \texttt{SNooPy}, are already capable of standardizing 1991bg-like SNe Ia.


\section{Conclusions}
\label{sec:discuss}

In this work, I have assembled a sample of 14 underluminous, 1991bg-like SNe Ia from the literature to test whether this type of SN Ia can be used as a standardizable candle, as previously noted by \citet{2018ApJ...869...56B} for fast-declining SNe Ia in general. To minimize systematic uncertainties, I have chosen SNe Ia that exploded in host galaxies with independent distance moduli, low to no reddening, and well-sampled light curves. Even so, the sample still suffers from systematic uncertainties that stem from the variety of filters in which the SNe were observed as well as a lack of $K$ and $S$ corrections. This sample still exhibits statistically significant ($>5\sigma$) correlations between the peak absolute magnitudes of the 1991bg-like SNe Ia and the widths of their light curves, as parameterized by the colour stretch, $s_\mathrm{BV}$. 

The 1991bg-like SN Ia width-luminosity relations measured here are similar in nature to those of normal SNe Ia but with significantly steeper slopes. The $B$- and $V$-band width-luminosity relations are consistent with preliminary CSP data, which both strengthens the existence of the correlations and shows that \texttt{SNooPy}, the light-curve fitter used by CSP, is able to standardize underluminous as well as normal SNe Ia. The correlations shown here are further strengthened by three test cases, namely SNe 1997cn, 1999da, and 2022xkq, which were not included in the calibration sample.

The work done here shows that 1991bg-like SNe Ia can be standardized and hence used to measure extragalactic distances. However, these width-luminosity relations must first be properly calibrated by compiling a homogeneous sample of 1991bg-like SNe Ia observed by a single survey in a single set of filters. Even then, because 1991bg-like SNe Ia are rarer than normal SNe Ia (15 vs. 70 per cent of all SNe Ia, respectively; \citealt{2017ApJ...837..121G}), their inclusion in SN Ia cosmology samples will not make much of a difference. Moreover, their inherent dimness makes them more susceptible to Malmquist bias, which limits their use as probes of dark energy. Finally, as shown here, the quadratic fits used by \citet{2018ApJ...869...56B} to simultaneously fit normal and fast-declining SNe Ia underestimate the absolute magnitudes of SNe Ia in the transition between normal and 1991bg-like SNe Ia, hinting that each of these subtypes should be fit separately.

Instead, I suggest using 1991bg-like SNe Ia as a rung on a new cosmological distance ladder, one that would provide an independent check on the distance ladder based on Cepheids and normal SNe Ia. While extremely successful (e.g., \citealt{2022ApJ...934L...7R}), the latter ladder is biased towards star-forming galaxies, which host young Cepheid variables. 1991bg-like SNe Ia, on the other hand, are mostly found in massive, passive galaxies \citep{2017ApJ...837..121G}. This makes them less prone to host-galaxy reddening and removes the systematic uncertainty produced by the mass step found in Hubble residuals (e.g., \citealt{Kelly2010,2021ApJ...909...26B,2021MNRAS.501.4861K,2023MNRAS.519.3046K,2023MNRAS.518.1985M}). 

Bereft of Cepheids, the galaxies that host 1991bg-like SNe Ia will have to be calibrated by other means, such as surface-brightness fluctuations. Several groups have already attempted to use this method to either measure $H_\mathrm{0}$ directly or to calibrate the host galaxies of normal SNe Ia (e.g., \citealt{2021ApJ...911...65B,2021ApJS..255...21J,2021A&A...647A..72K,2023ApJ...953...35G,2023arXiv230801875U,2023AAS...24142410W}). A new distance ladder, based on surface-brightness fluctuations (or some other non-Cepheid method) and 1991bg-like SNe Ia could provide an independent measurement of $H_\mathrm{0}$, a necessary step towards resolving the current `Hubble tension' (e.g., \citealt{2021CQGra..38o3001D,2023ARNPS..73..153K}).


\section*{Acknowledgments}
I thank Chris R. Burns and Saurabh W. Jha for helpful discussions; Willem B. Hoogendam, Jeniveve Pearson, and the CSP for sharing their data with me; and the anonymous referee for their comments. This research has made use of NASA's Astrophysics Data System and the NASA/IPAC Extragalactic Database (NED), which are funded by the National Aeronautics and Space Administration and operated by the California Institute of Technology. For the purpose of open access, the author has applied a Creative Commons Attribution (CC BY) license to any Author Accepted Manuscript version arising.

\section*{Data availability}
All data used in this work, except for the CSP data shown in Fig.~\ref{fig:hoogendam}, are public and have been published elsewhere. The measurements and fits produced by this work are included in the tables throughout this paper. Any remaining data or measurements will be shared on request to the corresponding author.


\appendix
\section{Distance modulus references}
\label{appendixA}

In this section, I summarize the sources of the various distance moduli measurements used to calculate the weighted means shown in Table~\ref{table:info}. The acronyms used in Table~\ref{table:dm}, below, are: Cepheids (Cep), cosmic microwave background (CMB), dwarf galaxy diameter (DGD), fundamental plane (FP), globular cluster luminosity function (GCLF), globular cluster radius (GCR), planetary nebula luminosity function (PNLF), surface brightness fluctuations (SBF), and the Tully-Fisher relation (TF).

\begin{table*}
 \caption{Distance modulus references}\label{table:dm}
 \begin{tabular}{llcl}
  \hline
  \hline
  SN & Host & Method & References \\
  \hline
  \multirow{6}{*}{1991bg}  & \multirow{6}{*}{NGC 4374} & FP        & \citet{2001MNRAS.327.1004B} \\
    & & GCLF  & \citet{2004AA...415..499G,2007ApJS..171..101J,2010ApJ...717..603V} \\
    & & GCR   & \citet{2005ApJ...634.1002J} \\
    & & PNLF  & \citet{2000ApJ...529..745F,2002ApJ...577...31C} \\
    & & SBF   & \citet{2000ApJ...529..745F,2001MNRAS.327.1004B,2009ApJ...694..556B,2007ApJ...655..144M,2011AA...532A.154C} \\
    & & TF        & \citet{2012ApJ...749..174C} \\        
  \hline
  1998de  & NGC 0252 & CMB & \citet{2001PASP..113..308M} \\
  \hline
  \multirow{2}{*}{1999by}  & \multirow{2}{*}{NGC 2841} & Cep & \citet{2001ApJ...559..243M,2006ApJS..165..108S} \\
    & & TF  & \citet{2002ApJ...565..681R,2009AJ....138..323T,2009ApJS..182..474S, 2012ApJ...749..174C,2012ApJ...758L..12S,2014MNRAS.444..527S} \\
  \hline
  \multirow{2}{*}{1999da}  & \multirow{2}{*}{NGC 6411} & FP  & \citet{2001MNRAS.327.1004B} \\
    & & SBF & \citet{2001MNRAS.327.1004B} \\
  \hline
  \multirow{2}{*}{2005bl}  & \multirow{2}{*}{NGC 4059} & FP & \citet{2016AA...596A..14S} \\
    & & TF & \citet{2007AA...465...71T} \\
  \hline
  \multirow{8}{*}{2006mr}  & \multirow{8}{*}{NGC 1316} & $D$--$\sigma$ & \citet{2002AJ....123.2159B} \\
    & & DGD  & \citet{2011MNRAS.414.3699M} \\
    & & FP   & \citet{2002MNRAS.334..859G} \\
    & & GCLF & \citet{2010ApJ...717..603V} \\
    & & GCR  & \citet{2010ApJ...715.1419M} \\
    & & PNLF & \citet{2002ApJ...577...31C,2007ApJ...657...76F} \\
    & & \multirow{2}{*}{TF}  & \citet{2000ApJ...529..698S,2000ApJ...533..744T,2007AA...465...71T,2012ApJ...749...78T} \\
    & & & \citet{2012ApJ...749..174C,2013ApJ...765...94S,2014MNRAS.444..527S,2014ApJ...792..129N,2016AJ....152...50T} \\
    & & \multirow{2}{*}{SBF} & \citet{2001ApJ...546..681T,2001ApJ...559..584A,2003ApJ...583..712J,2007ApJ...668..130C} \\
    & & & \citet{2009ApJ...694..556B,2010ApJ...724..657B,2013AA...552A.106C} \\
  \hline
  2007ax  & NGC 2577 & TF  & \citet{2007AA...465...71T} \\
  \hline
  2008R   & NGC 1200 & FP  & \citet{2014MNRAS.445.2677S} \\
  \hline
  \multirow{2}{*}{2015bo}  & \multirow{2}{*}{NGC 5490} & TF  & \citet{2007AA...465...71T} \\
    & & SBF  & \citet{2021ApJS..255...21J} \\
  \hline
  \multirow{6}{*}{2017ejb} & \multirow{6}{*}{NGC 4696} & DGD & \citet{2011MNRAS.414.3699M} \\
    & & FP     & \citet{2001MNRAS.327.1004B} \\
    & & FP+SBF & \citet{2013AJ....146...86T} \\
    & & GCLF & \citet{2005AA...438..103M} \\
    & & TF  & \citet{2007AA...465...71T,2016AJ....152...50T} \\
    & & SBF & \citet{2001ApJ...546..681T,2001MNRAS.327.1004B,2003AA...410..445M,2005ApJ...634..239C} \\
  \hline
  2019so  & NGC 4622 & FP & \citet{2014MNRAS.445.2677S} \\
  \hline
  2021qvv & NGC 4442 & GCLF & \citet{2007ApJS..171..101J,2010ApJ...717..603V} \\
    & & GCR & \citet{2005ApJ...634.1002J} \\
    & & TF  & \citet{2007AA...465...71T} \\
    & & SBF & \citet{2007ApJ...655..144M,2009ApJ...694..556B} \\
  \hline
  \multirow{2}{*}{2022xkq} & \multirow{2}{*}{NGC 1784} & \multirow{2}{*}{TF} & \citet{2007AA...465...71T,2009AJ....138..323T,2011AA...532A.104N,2013AJ....146...86T,2016AJ....152...50T,2013ApJ...771...88L} \\
   & & & \citet{2014MNRAS.444..527S} \\
  \hline
  \multicolumn{4}{l}{\textbf{Note.} SN 1997cn exploded in the same host galaxy as SN 2015bo, NGC 5490.}
 \end{tabular}
\end{table*}

\end{document}